\documentclass[aps,preprint]{revtex4}
\usepackage{amssymb,epsf}
\usepackage{amsmath}
\usepackage{graphicx}

\begin{document}

\title{Thermodynamics of Rotating Solutions in
 Gauss-Bonnet-Maxwell Gravity and the Counterterm Method}
\author{M. H. Dehghani$^{1,2}$\footnote{email address:
mhd@shirazu.ac.ir} G. H. Bordbar$^{1,2}$\footnote{email address:
bordbar@physics.susc.ac.ir} and M. Shamirzaie$^{1}$}
\affiliation{$^1$Physics Department and Biruni Observatory,
College of Sciences, Shiraz
University, Shiraz 71454, Iran\\
$^2$Research Institute for Astrophysics and Astronomy of Maragha
(RIAAM), Maragha, Iran}
\begin{abstract}
By a suitable transformation, we present the $(n+1)$-dimensional
charged rotating solutions of Gauss-Bonnet gravity with a complete
set of allowed rotation parameters which are real in the whole
spacetime. We show that these charged rotating solutions present
black hole solutions with two inner and outer event horizons,
extreme black holes or naked singularities provided the parameters
of the solutions are chosen suitable. Using the surface terms that
make the action well-defined for Gauss-Bonnet gravity and the
counterterm method for eliminating the divergences in action, we
compute finite action of the solutions. We compute the conserved
and thermodynamical quantities through the use of free energy and
the counterterm method, and find that the two methods give the
same results. We also find that these quantities satisfy the first
law of thermodynamics. Finally, we perform a stability analysis by
computing the heat capacity and the determinant of Hessian matrix
of mass with respect to its thermodynamic variables in both the
canonical and the grand-canonical ensembles, and show that the
system is thermally stable. This is commensurate with the fact
that there is no Hawking-Page phase transition for black objects
with zero curvature horizon.

\end{abstract}
%\pacs{04.50.+h, 04.20.Jb, 04.70.Bw, 04.70.Dy}
\maketitle

\section{Introduction}

\label{Intro} The thermodynamics of asymptotically anti-de Sitter
(AdS) black holes, which produces an aggregate of ideas from
thermodynamics, quantum field theory and classical gravity,
continues to attract a great deal of attention. One reason for
this is the role of AdS/CFT duality \cite {Mal1}, and in
particular, Witten interpretation of the Hawking-Page phase
transition between thermal AdS and asymptotically AdS black hole
as the confinement-deconfinment phases of the Yang-Mills theory
defined on the asymptotic boundaries of the AdS geometry
\cite{Wit1}. The fact that thermodynamics of black holes may be
modified by using other theories of gravity with higher derivative
(HD) curvature terms provides a strong motivation for considering
thermodynamics of black holes in HD gravity \cite{Ish}. The
appearance of HD gravitational terms can be seen, for example, in
the renormalization of quantum field theory in curved spacetime
\cite{BDav}, or in the construction of low-energy effective action
of string theory \cite {Wit2}. Among the theories of gravity with
higher derivative curvature terms, the Gauss-Bonnet gravity has
some special features in some sense. Indeed, in order to have a
ghost-free action, the quadratic curvature corrections to the
Einstein-Hilbert action should be proportional to the Gauss-Bonnet
term \cite{Zw,Des}. This combination also appear naturally in the
next-to-leading order term of the heterotic string effective
action, and plays a fundamental role in Chern-Simons gravitational
theories \cite{Cham}. From a geometric point of view, the
combination of the Einstein-Gauss-Bonnet terms constitutes, for
five dimensional spacetimes, the most general Lagrangian producing
second order field equations, as in the four-dimensional gravity
which the Einstein-Hilbert action is the most general Lagrangian
producing second order field equations \cite{Lov}.

To investigate the thermodynamics of black holes, one should compute the
finite conserved and thermodynamic quantities of the system. In general the
gravitational action is divergent when evaluated on solutions, as is the
Hamiltonian and other associated conserved quantities. A common approach to
evaluating thermodynamic quantities has been to carry out all computations
relative to some other spacetime that is regarded as the ground state for
the class of spacetimes of interest. This could be done by taking the
original action for gravity coupled to matter fields and subtracting from it
a reference action, which is a functional of the induced metric $\gamma $ on
the boundary $\partial M$ \cite{BY}. Conserved and/or thermodynamic
quantities are then computed relative to this boundary, which can then be
taken to infinity if desired. Unfortunately, it suffers from several
drawbacks. The choice of reference spacetime is not always unique \cite{CCM}%
, nor is it always possible to embed a boundary with a given
induced metric into the reference background. An extension of this
approach which addresses these difficulties was developed based on
the conjectured AdS/CFT correspondence for asymptotic AdS
spacetimes \cite{BK,EJM}. Applications of AdS/CFT correspondence
also include computations of conserved quantities for black holes
with rotation, NUT charge, various topologies, rotating black
strings with zero curvature horizons and rotating higher genus
black branes \cite{Deh-topologic,Deh,Od1}. Although the
counterterm method applies for the case of a specially infinite
boundary, it was also employed for the computation of the
conserved and thermodynamic quantities in the case of a
finite boundary \cite{DM2}. It is believed that appending a counterterm, $I_{%
\mathrm{ct}}$, to the action which depends only on the intrinsic
geometry of the boundary(ies) can remove the divergences. This
requirement, along with general covariance, implies that these
terms are functionals of curvature invariants of the induced
metric and have no dependence on the extrinsic curvature of the
boundary(ies)\cite{EJM}. An algorithmic procedure exists for
constructing $I_{\mathrm{ct}}$ for asymptotic AdS in Einstein
gravity and so its determination in this theory is unique
\cite{kls}. But in HD gravity these conterterms have been not
introduced till now. Since these counterterms should be functions
of intrinsic curvature of boundary, the construction of these
terms for spacetimes with flat boundaries is straightforward
\cite{DM,DBS}.

The outline of our paper is as follows. We give a brief review of the field
equations of Gauss-Bonnet gravity and the counterterm method for calculating
conserved quantities in Sec. \ref{conserved}. In Sec. \ref{Therm} we obtain
mass, angular momentum, entropy, temperature, charge, and electric potential
of the $(n+1)$-dimensional black hole solutions with a complete set of
rotational parameters and show that these quantities satisfy the first law
of thermodynamics. We also perform a local stability analysis of the black
holes in the canonical and grand canonical ensembles. We finish our paper
with some concluding remarks.

\section{The action and conserved quantities}

\label{conserved} The gravitational action of an $(n+1)$-dimensional
asymptotically anti-de Sitter spacetimes $\mathcal{M}$ with the Gauss-Bonnet
term in the present of an electromagnetic field is:
\begin{equation}
\begin{aligned} I_{G} &=-\frac{1}{16\pi
}\int_{\mathcal{M}}d^{n+1}x\sqrt{-g}\{R-2\Lambda +\alpha \left( R_{\mu \nu
\sigma \kappa }R^{\mu \nu \sigma \kappa }-4R_{\mu \nu }R^{\mu \nu
}+R^{2}\right) -F_{\mu \nu }F^{\mu \nu }\} \\ &\quad-\frac{1}{8\pi
}\int_{\partial \mathcal{M}}d^{n}x\sqrt{-\gamma }\left\{ \Theta +2\alpha
\left( J-2\mathcal{G}_{ab}\Theta ^{ab}\right) \right\} \end{aligned}
\label{act1}
\end{equation}
where $F_{\mu \nu }=\partial _{\mu }A_{\nu }-\partial _{\nu
}A_{\mu }$ is electromagnetic tensor field, $A_{\mu }$ is the
vector potential, $\Lambda =-n(n-1)/2l^{2}$ is the cosmological
constant, $\alpha $ is Gauss-Bonnet coefficient with dimensions
$(\mathrm{length})^{2}$ which we assume to be
positive as in heterotic string theory, $R_{\mu \nu \sigma \kappa }$, $%
R_{\mu \nu }$ and $R$ are Riemann tensor, Ricci tensor and Ricci
scalar of the manifold $\mathcal{M}$ respectively, $\gamma _{ab}$
is induced metric on the boundary $\partial \mathcal{M}$, $\Theta
$ is trace of extrinsic curvature of the boundary,
$\mathcal{G}_{ab}(\gamma )$ is Einstein tensor calculated on the
boundary, and $J$ is trace of:
\begin{equation}
J_{ab}=\frac{1}{3}(\Theta _{cd}\Theta ^{cd}\Theta _{ab}+2\Theta \Theta
_{ac}\Theta _{b}^{c}-2\Theta _{ac}\Theta ^{cd}\Theta _{db}-\Theta ^{2}\Theta
_{ab})  \label{psi}
\end{equation}
The second integral is a surface term which is chosen such that the
variational principle is well-defined \cite{Myers-surf,Dav-surf}. In order
to obtain the field equations by the variation of the volume integral with
respect to the fields, one should impose the boundary condition $\delta
A_{\mu }=0$ on $\partial \mathcal{M}$. Thus the action (\ref{act1}) is
appropriate to study the grand-canonical ensemble with fixed electric
potential \cite{Cal}. To study the canonical ensemble with fixed electric
charge one should impose the boundary condition $\delta (n^{a}F_{ab})=0$,
and therefore the gravitational action is \cite{Haw-Ross}:
\begin{equation}
\overset{\sim }{I}_{G}=I_{G}-\frac{1}{4\pi }\int_{\partial \mathcal{M}%
_{\infty }}d^{n}x\sqrt{_{-}\gamma }n_{a}F^{ab}A_{b},  \label{act2}
\end{equation}
where $n_{a}$\ is the normal to the boundary $\partial \mathcal{M}$. Varying
the action (\ref{act1}) or (\ref{act2}) with respect to the metric tensor $%
g_{\mu \nu }$ and electromagnetic tensor field $F_{\mu \nu }$, with
appropriate boundary condition, the equations of gravitation and
electromagnetic fields are obtained as
\begin{eqnarray}  \label{field}
&&G_{\mu \nu }+\Lambda g_{\mu \nu }+\alpha G_{\mu \nu }^{(2)}=T_{\mu \nu }
\label{fiel1} \\
&&\nabla _{\nu }F^{\mu \nu }=0
\end{eqnarray}
where $G_{\mu \nu }$ is the Einstein tensor, $T_{\mu \nu }^{\mathrm{(em)}%
}=2F_{\phantom{\lambda}{\mu}}^{\rho }F_{\rho \nu }-\frac{1}{2}F_{\rho \sigma
}F^{\rho \sigma }g_{\mu \nu }$ is the energy-momentum tensor of
electromagnetic field and $G_{\mu \nu }^{(2)}$ is second order Lovelock
tensor:
\begin{eqnarray}
G_{\mu \nu }^{(2)} &=&2(R_{\mu \sigma \kappa \tau }R_{\nu }^{\phantom{\nu}%
\sigma \kappa \tau }-2R_{\mu \rho \nu \sigma }R^{\rho \sigma }-2R_{\mu
\sigma }R_{\phantom{\sigma}\nu }^{\sigma }+RR_{\mu \nu })  \notag
\label{Gaussfield} \\
&&-\frac{1}{2}\left( R_{\mu \nu \sigma \kappa }R^{\mu \nu \sigma \kappa
}-4R_{\mu \nu }R^{\mu \nu }+R^{2}\right) g_{\mu \nu }
\end{eqnarray}
In general the action $I_{G}$, is divergent when evaluated on the solutions,
as is the Hamiltonian and other associated conserved quantities. Rather than
eliminating these divergences by incorporating reference term, a counterterm
$I_{\mathrm{ct}}$ is added to the action which is functional only of the
boundary curvature invariants. For asymptotically AdS spacetimes with flat
boundary, all the counterterm containing the curvature invariants of the
boundary are zero and therefore the counterterm reduces to a volume term as
\begin{equation}
I_{\mathrm{ct}}=-\frac{(n-1)}{8\pi L}\int_{\partial \mathcal{M}}d^{n}x\sqrt{%
-\gamma }\   \label{Ict}
\end{equation}
where $L$ is a length factor that depends on the fundamental constants $l$
and $\alpha $, that must reduce to $l$ as $\alpha $ goes to zero. Thus, the
total action, $I$, can be written as
\begin{equation}
I=I_{G}+I_{\mathrm{ct}}  \label{Itot}
\end{equation}
Having the total finite action, one can use Brown and York definition \cite
{BY} to construct a divergence free stress-energy tensor as
\begin{mathletters}
\begin{equation}
T^{ab}=\frac{1}{\kappa }\left\{ \left( \Theta ^{ab}-\Theta \gamma
^{ab}\right) +2\alpha \left( 3J^{ab}-J\gamma ^{ab}\right) +\frac{n-1}{L}%
\gamma ^{ab}\right\}  \label{Tab}
\end{equation}
To compute the conserved charges of the spacetime, one should choose a
spacelike hypersurface $\mathcal{B}$ in $\partial \mathcal{M}$ with metric $%
\sigma _{ij}$, and write the boundary metric in ADM form:
\end{mathletters}
\begin{equation*}
\gamma _{ab}dx^{a}dx^{a}=-N^{2}dt^{2}+\sigma _{ij}\left( d\phi
^{i}+V^{i}dt\right) \left( d\phi ^{j}+V^{j}dt\right) ,
\end{equation*}
where the coordinates $\phi ^{i}$ are the angular variables parameterizing
the hypersurface of constant $r$ around the origin. When there is a Killing
vector field $\mathcal{\xi }$ on the boundary, then the conserved quantities
associated with the stress tensors of Eq. (\ref{Tab}) can be written as
\begin{equation}
\mathcal{Q}(\mathcal{\xi )}=\int_{\mathcal{B}_{\infty }}d^{n-1}\phi \sqrt{%
\sigma }T_{ab}n^{a}\mathcal{\xi }^{b},  \label{charge}
\end{equation}
where $\sigma $ is the determinant of the metric $\sigma _{ij}$ and $n^{a}$
is the unit normal vector to the boundary $\mathcal{B}$ . For boundaries
with timelike ($\xi =\partial /\partial t$) and rotational Killing vector
fields ($\zeta =\partial /\partial \phi $), we obtain
\begin{eqnarray}  \label{J}
M &=&\int_{\mathcal{B}_{\infty }}d^{n-1}\phi \sqrt{\sigma }T_{ab}n^{a}\xi
^{b},  \label{Mastot} \\
J &=&\int_{\mathcal{B}_{\infty }}d^{n-1}\phi \sqrt{\sigma }T_{ab}n^{a}\zeta
^{b},  \label{Angtot}
\end{eqnarray}
provided the surface $\mathcal{B}$ contains the orbits of $\zeta $. These
quantities are, respectively, the conserved mass and angular momentum of the
system enclosed by the boundary. Note that they will both be dependent on
the location of the boundary $\mathcal{B}$ in the spacetime, although each
is independent of the particular choice of foliation $\mathcal{B}$ within
the boundary $\partial \mathcal{M}$. In the context of the (A)dS/CFT
correspondence, the limit in which the boundary $\mathcal{B}$ becomes
infinite ($\mathcal{B}_{\infty }$) is taken, and the counterterm
prescription ensures that the action and conserved charges are finite. No
embedding of the surface $\mathcal{B}$ into a reference spacetime is
required and the quantities which are computed are intrinsic to the
spacetimes.

\section{Thermodynamics of black branes \label{Therm}}

\label{metric}The rotation group in $n+1$ dimensions is $SO(n)$ and
therefore the number of independent rotation parameters for a localized
object is equal to the number of Casimir operators, which is $[n/2]\equiv k$%
, where $[n/2]$ is the integer part of $n/2$. Asymptotically AdS solution of
the field equations (\ref{field}) with cylindrical symmetry with $k$
rotation parameter $a_{i}$, can be written as \cite{Deh4}:
\begin{equation*}
\begin{split}
ds^{2}& =-f(\rho )\left( \Xi dt-{{\sum_{i=1}^{k}}}a_{i}d\phi _{i}\right)
^{2}+\frac{\rho ^{2}}{l^{4}}{{\sum_{i=1}^{k}}}\left( a_{i}dt-\Xi l^{2}d\phi
_{i}\right) ^{2} \\
& -\frac{\rho ^{2}}{l^{2}}{\sum_{i=1}^{k}}(a_{i}d\phi _{j}-a_{j}d\phi
_{i})^{2}+\frac{d\rho ^{2}}{f(\rho )}+\rho ^{2}dX^{2}, \\
\Xi ^{2}& ={{\sum_{i=1}^{k}}}\left( 1+\frac{a_{i}^{2}}{l^{2}}\right) , \\
A_{\mu }& =\sqrt{\frac{(n-1)}{2(n-2)}}\frac{q}{\rho ^{n-2}}(\Xi \delta _{\mu
}^{0}-a_{i}\delta _{\mu }^{i}),\hspace{0.5cm}\text{(no sum on }i\text{).}
\end{split}\label{met1}
\end{equation*}
where $dX^{2}$ is the Euclidean metric on the $(n-k-1)$-dimensional
submanifold with volume $\omega _{n-k-1}$ and $f(r)$ is
\begin{equation}
f(\rho )=\frac{\rho ^{2}}{2(n-2)(n-3)\alpha }\left\{ 1-\sqrt{%
1-4(n-2)(n-3)\alpha \left( \frac{1}{l^{2}}-\frac{m}{\rho ^{n}}+\frac{q^{2}}{%
\rho ^{2n-2}}\right) }\right\} ,  \label{Fr1}
\end{equation}
As we will see in the next section, $m$ and $q$ are related to the total
mass and total charge of spacetimes. Although $f(\rho )$ for the uncharged
solution ($q=0$) is real in the whole range $0\leq \rho <\infty $ provided $%
\alpha \leq l^{2}/4(n-2)(n-3)$, for charged solution it is real only in the
range $r_{0}\leq \rho <\infty $ where $r_{0}$ is the largest real root of
the following equation:
\begin{equation}
\left( \frac{1}{4(n-2)(n-3)}-\frac{\alpha }{l^{2}}\right)
r_{0}{}^{(2n-2)}+\alpha mr_{0}{}^{(n-2)}-\alpha q^{2}=0  \label{r0}
\end{equation}
One may note that the only solution of Eq. (\ref{r0}) for the case of
uncharged solution is $r_{0}=0$. In order to restrict the spacetime to the
region $\rho \geq r_{0}$, we introduce a new radial coordinate $r$ as:
\begin{equation*}
r^{2}=\rho ^{2}-r_{0}^{2}\Rightarrow d\rho ^{2}=\frac{r^{2}}{r^{2}+r_{0}^{2}}%
dr^{2}.
\end{equation*}
With this new coordinate, the above metric becomes:
\begin{eqnarray}
ds^{2} &=&-f(r)\left( \Xi dt-{{\sum_{i=1}^{k}}}a_{i}d\phi _{i}\right) ^{2}+%
\frac{(r^{2}+r_{0}^{2})}{l^{4}}{{\sum_{i=1}^{k}}}\left( a_{i}dt-\Xi
l^{2}d\phi _{i}\right) ^{2}  \notag \\
&&-\frac{r^{2}+r_{0}^{2}}{l^{2}}{\sum_{i=1}^{k}}(a_{i}d\phi _{j}-a_{j}d\phi
_{i})^{2}+\frac{r^{2}dr^{2}}{(r^{2}+r_{0}^{2})f(r)}+(r^{2}+r_{0}^{2})d\Omega
^{2},  \label{met2}
\end{eqnarray}
where now the gauge potential and the metric function are:
\begin{eqnarray}
&& A_{\mu } =\sqrt{\frac{(n-1)}{2(n-2)}}\frac{q}{(r^{2}+r_{0}^{2})^{\frac{n}{2}%
-1}}((\Xi \delta _{\mu }^{0}-a_{i}\delta _{\mu }^{i}),\hspace{1cm}\text{(no
sum on }i\text{).}  \label{A2} \\
&& f(r) =\frac{r^{2}+r_{0}^{2}}{2(n-2)(n-3)\alpha }\left\{ 1-\sqrt{%
1-4(n-2)(n-3)\left( \frac{\alpha }{l^{2}}-\frac{\alpha m}{(r^{2}+r_{0}^{2})^{%
\frac{n}{2}}}+\frac{\alpha q^{2}}{(r^{2}+r_{0}^{2})^{n-1}}\right)
}\right\} \label{fr2}
\end{eqnarray}
It is notable to mention that the in the static case $a_{i}=0$, in
contrast with the case of uncharged solution, $g_{tt}$ is not
equal to $g_{rr}^{-1}$.

\subsection{Properties of the solutions}\label{property}
One can show that the Kretschmann scalar $R_{\mu
\nu \lambda \kappa }R^{\mu \nu \lambda \kappa }$ diverges at
$r=0$, and therefore there is a curvature singularity located at
$r=0$. Seeking possible black hole solutions, we turn to looking
for the existence of horizons. As in the case of rotating black
hole solutions of Einstein
gravity, the above metric given by Eqs. (\ref{met2})-(\ref{fr2}%
) has both Killing and event horizons. The Killing horizon is a null surface
whose null generators are tangent to a Killing field. The proof that a
stationary black hole event horizon must be a Killing horizon in the
four-dimensional Einstein gravity \cite{Haw1+3} cannot obviously be
generalized to higher order gravity. However the result is true for all
known static solutions. Although our solution is not static, the Killing
vector
\begin{equation}
\chi =\partial _{t}+{{{\sum_{i=1}^{k}}}}\Omega _{i}\partial _{\phi _{i}},
\label{Kil}
\end{equation}
is the null generator of the event horizon, where $k$ denotes the number of
rotation parameters. The event horizon define by the solution of $%
g^{rr}=f(r)=0$. The metric of Eqs. (\ref{met2})-(\ref{fr2}) has
two inner and outer event horizons located at $r_{-}$ and $r_{+}$,
provided the mass parameter $m$ is greater than $m_{\mathrm{ext}}$
given as:
\begin{equation}
m_{\mathrm{ext}}=2\left( \frac{n-1}{n-2}\right) \left(
\frac{n}{n-2}\right)
^{-\frac{n-2}{2(n-1)}}l^{-\frac{n-2}{n-1}}q^{\frac{n}{n-1}}\label{mext}.
\end{equation}
In the case that $m=m_{\mathrm{ext}}$, we will have an extreme black brane.
One can obtain the angular velocity of the event horizon by analytic
continuation of the metric. Setting $a_{i}\rightarrow ia_{i}$ yields the
Euclidean section of (\ref{met2}), whose regularity at $r=r_{+}$ requires
that we should identify $\phi _{i}\sim \phi _{i}+\beta _{+}\Omega _{i}$,
where $\Omega _{i}$'s are the angular velocities of the outer event horizon.
One obtains:
\begin{equation}
\Omega _{i}=\frac{a_{i}}{\Xi l^{2}}\label{Om1}.
\end{equation}
The temperature may be obtained through the use of definition of surface
gravity,
\begin{equation}
T_{+}=\frac{1}{2\pi }\sqrt{-\frac{1}{2}\left( \nabla _{\mu }\chi _{\nu
}\right) \left( \nabla ^{\mu }\chi ^{\nu }\right) ,}
\end{equation}
where $\chi $ is the Killing vector (\ref{Kil}). One obtains:
\begin{equation}
\beta^{-1}=T_{+}=\frac{f^{\prime }(r_{+})}{4\pi \Xi }\sqrt{1+\frac{r_{0}^{2}}{r_{+}^{2}}%
}=\frac{n(r_{+}^{2}+r_{0}^{2})^{n-1}-(n-2)l^{2}q^{2}}{4\pi \Xi
l^{2}r_{+}(r_{+}^{2}+r_{0}^{2})^{n-2}}  \label{Temp}
\end{equation}
One may note that the temperature (\ref{Temp}) reduces to the temperature of
the uncharged solution for the uncharged solutions where $r_{0}=0$ \cite
{Deh4}. Next, we calculate the electric charge of the solutions. To
determine the electric field we should consider the projections of the
electromagnetic field tensor on special hypersurfaces. The normal to such
hypersurfaces is
\begin{equation}
u^{0}=\frac{1}{N},\qquad u^{r}=0,\qquad u^{i}=\frac{V^{i}}{N},
\end{equation}
and the electric field is $E^{\mu }=g^{\mu \rho }F_{\rho \nu }u^{\nu }$,
where $N$ and $V^{i}$ are the lapse and shift functions. Then the electric
charge per unit volume $V_{n-1}$ can be found by calculating the flux of the
electric field at infinity, yielding
\begin{equation}
Q=\frac{\Xi }{4\pi }\sqrt{\frac{(n-1)(n-2)}{2}}q.  \label{Ch}
\end{equation}
The electric potential $\Phi $, measured at infinity with respect to the
horizon, is defined by \cite{Gub,Cal}:
\begin{equation}
\Phi =A_{\mu }\chi ^{\mu }\left| _{r\rightarrow \infty }-A_{\mu }\chi ^{\mu
}\right| _{r=r_{+}},  \label{Pot1}
\end{equation}
where $\chi $ is the null generator of the horizon given by Eq. (\ref{Kil}).
We find
\begin{equation}
\Phi =\sqrt{\frac{(n-1)}{2(n-2)}}\frac{q}{\Xi (r_{+}^{2}+r_{0}^{2})^{\frac{%
(n-2)}{2}}}.  \label{Pot2}
\end{equation}

\subsection{Action and conserved quantities}

\label{action-sub} The action and conserved quantities associated with the
spacetime described by (\ref{met2}) can be obtained via the counterterm
method. Using Eqs. (\ref{act1}), (\ref{act2}) and (\ref{Ict}), the Euclidean
actions will be finite provided
\begin{equation}
\frac{1}{L}=\tfrac{1}{3\sqrt{2(n-2)(n-3)}}\tfrac{l^{2}+4(n-2)(n-3)\,\alpha -l%
\sqrt{l^{2}-4(n-2)(n-3)\,\alpha }}{l\;\sqrt{\alpha l\left( l-\sqrt{%
l^{2}-4(n-2)(n-3)\alpha }\right) }}
\end{equation}
where we note that $L$ reduces to $l$ as $\alpha $ goes to zero. Denoting
the volume of the hypersurface boundary at constant $t$ and $r$ by $%
V_{n-1}=(2\pi )^{k}\omega _{n-k-1}$, the Euclidean actions per unit volume $%
V_{n-1}$ in canonical and grand-canonical ensembles can then be obtained
through the use of Eqs. (\ref{act1}), (\ref{act2}) and (\ref{Ict}). We find:
\begin{align}
I_{n}& =-\frac{\beta _{+}}{16\pi }\frac{(r_{+}^{2}+r_{0}^{2})^{n-1}+q^{2}%
\;l^{2}}{l^{2}\,(r_{+}^{2}+r_{0}^{2})^{\frac{n}{2}-1}}, \\
\tilde{I}_{n}& =-\frac{\beta _{+}}{16\pi }\frac{%
(r_{+}^{2}+r_{0}^{2})^{n-1}-(2n-3)q^{2}\;l^{2}}{l^{2}%
\,(r_{+}^{2}+r_{0}^{2})^{\frac{n}{2}-1}}
\end{align}

We first obtain the action in grand canonical ensemble as a
function of the intensive quantities $\beta $, $\mathbf{\Omega }$
and $\Phi$ by using the expression for the temperature, the
angular velocity and the potential given in Eqs. (\ref{Temp}),
(\ref{Om1}) and (\ref{Pot2}):
\begin{equation}
I=-\frac{\beta (r_{+}^{2}+r_{0}^{2})^{n-2}}{16\pi
l^{2}\,(r_{+}^{2}+r_{0}^{2})^{\frac{n}{2}-1}}\left( (r_{+}^{2}+r_{0}^{2})+%
\frac{2(n-2)l^{2}}{(n-1)(1-\Omega ^{2}l^{2})}\right) ,  \label{Smar}
\end{equation}
where $r_{+}$  in terms of $\beta $, $\mathbf{\Omega }$ and $\Phi$
is
\begin{equation}
r_{+}=\frac{\pi (n-1)+\sqrt{\left[ \pi ^{2}(n-1)^{2}-\Lambda
\Phi^{2}\beta ^{2}(n-2)^{2}\right] }}{\beta \Lambda \sqrt{1-\Omega
^{2}l^{2}}}.
\end{equation}

Since the Euclidean action is related to the free energy in the grand
canonical ensemble, the electric charge $Q$, the angular momenta $J_{i}$,
the entropy $S$ and the mass $M$ per unit volume $V_{n-1}$ can be found
using the familiar thermodynamics relations:
\begin{eqnarray}
Q &=&-\beta ^{-1}\frac{\partial I}{\partial \Phi}=\frac{\Xi }{4\pi }\sqrt{\frac{%
(n-1)(n-2)}{2}}q,  \label{qq} \\
S &=&\left( \beta \frac{\partial }{\partial \beta }-1\right) I=\frac{\Xi }{4}%
(r_{+}^{2}+r_{0}^{2})^{\frac{n-1}{2}},  \label{ent} \\
J_{i} &=&-\beta ^{-1}\frac{\partial I}{\partial \Omega _{i}}=\frac{1}{16\pi }%
n\Xi ma_{i},  \label{Jtot1} \\
M &=&\left( \frac{\partial }{\partial \beta }-\beta ^{-1}\Phi \frac{\partial }{%
\partial \Phi}-\beta ^{-1}\sum \Omega _{i}\frac{\partial }{\partial \Omega _{i}}%
\right) I=\frac{1}{16\pi }(n\Xi ^{2}-1)m  \label{Mtot1}
\end{eqnarray}
Note that the charged obtained by use of free energy is the same
as that obtained in Eq. (\ref{Ch}).
Of course, if one computes the mass and angular momentum per unit volume $%
V_{n-1}$ through the use of counterterm method introduced in Ref.
\cite{DM} (explained in Sec. \ref {conserved}), the same results
will be obtained.

Equation (\ref{ent}) shows that the entropy of the black brane satisfies the
so-called area law of entropy which states that the black hole entropy
equals to one-quarter of horizon area \cite{Beck}. Although the area law of
entropy applies to almost all kind of black holes and black strings of
Einstein gravity \cite{Haw222}, it is not satisfied in higher derivative
gravity \cite{fails+1}. But, for our solutions in Gauss-Bonnet gravity with
flat horizon the area law is satisfied.

\subsection{Stability of black branes}

\label{stability} We first obtain the mass $M$ as a function of
$S$, $Q$ and $\mathbf{J}$. Using the expressions (\ref{ent}),
(\ref{Mtot1}) and (\ref {Jtot1}) for the entropy, the mass and the
angular momentum, and the fact that $f(r_{+})=0$, one obtains by
simple algebraic manipulation the Smarr-type formula as:
\begin{equation}
M(S,J,Q)=\frac{(nZ-1)J}{nl\sqrt{Z(Z-1)}},
\end{equation}
where $J^{2}=\left| \mathbf{J}\right| ^{2}=\sum_{i}^{k}J_{i}^{2}$ and $Z=\Xi
^{2}$ is the real positive solution of the following equation:
\begin{equation}
\left( Z-1\right) ^{(d-2)}-\frac{Z}{16S^{2}}\left\{ \frac{4\pi (n-1)(n-2)lSJ%
}{n[(n-1)(n-2)S^{2}+2\pi ^{2}Q^{2}l^{2}]}\right\} ^{(2n-2)}=0.
\end{equation}
It is worthwhile to mention that the thermodynamic quantities calculated
above satisfy the first law of thermodynamics,
\begin{equation}
dM=TdS+\sum_{i}\Omega _{i}dJ_{i}+\Phi dQ.  \label{1law}
\end{equation}
The stability of a thermodynamic system with respect to the small variations
of the thermodynamic coordinates, is usually performed by analyzing the
behavior of the entropy $S(M,Q,\mathbf{J})$ around the equilibrium. The
local stability in any ensemble requires that $S(M,Q,\mathbf{J})$ be a
convex function of their extensive variables or its Legendre transformation
must be a concave function of their intensive variables. Thus, the local
stability can in principle be carried out by finding the determinant of the
Hessian matrix of $S$ with respect to its extensive variables, $\mathbf{H}%
_{X_{i}X_{j}}^{S}=[\partial ^{2}S/\partial X_{i}\partial X_{j}]$, where $%
X_{i}$'s are the thermodynamic variables of the system. Indeed, the system
is locally stable if the determinant of Hessian matrix satisfies $\mathbf{H}%
_{X_{i},X_{j}}^{S}\leq 0$ \cite{Gub,Cal}. Also, one can perform the
stability analysis through the use of the determinant of Hessian matrix of
the energy with respect to its thermodynamic variables, and the stability
requirement $\mathbf{H}_{X_{i},X_{j}}^{S}\leq 0$ may be rephrased as $%
\mathbf{H}_{Y_{i},Y_{j}}^{M}\geq 0$ \cite{Gus}.

\begin{figure}[tbp]
\epsfxsize=8cm \centerline{\epsffile{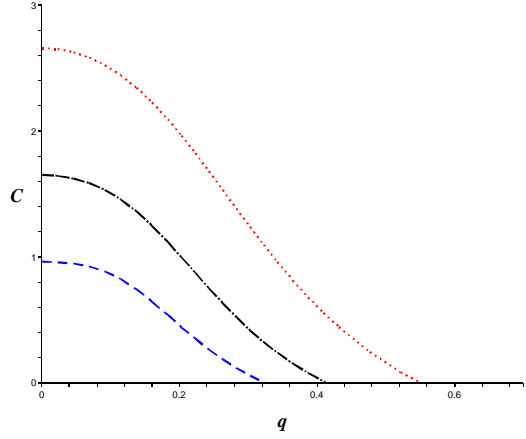}}
\caption{$C_{J,Q}$ versus $q$ for $l=12$, $r_+=1.4$, $n=4$ (dashed), $n=5$
(dash-dotted), and $n=6$ (dotted).}
\label{Figure1}
\end{figure}

\begin{figure}[tbp]
\centerline{\includegraphics[height=8cm,angle=0,width=8cm]{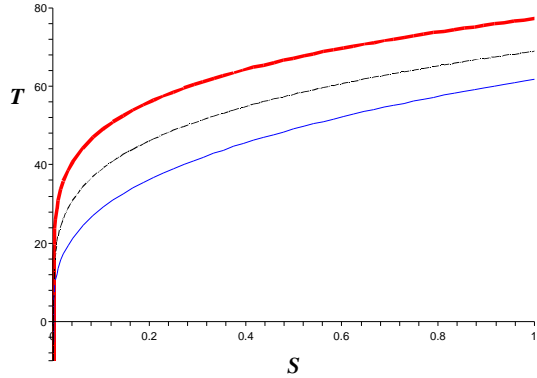}}
\caption{$T$ versus $S$ for $l=0.09$, $q=0.06$, $a=0.01$, $n=4$ (solid
line)), $n=5$ (dash-dotted), and $n=6$ (bold line).}
\label{Figure2}
\end{figure}

The number of the thermodynamic variables depends on the ensemble which is
used. In the canonical ensemble, the charge and angular momenta are fixed
parameters, and therefore the positivity of the heat capacity $C_{Q,\mathbf{J%
}}=T(\partial S/\partial T)_{Q,\mathbf{J}}$ is sufficient to assure the
local stability. One can show that the latter criteria is: \bigskip
\begin{equation}  \label{capacity}
\begin{aligned} C_{\mathbf{J},Q} &=\frac{\Xi
}{4}\left(r_{0}^{2}+r_{+}^{2}\right)^{\frac{(n-1)}{2}}\Bigg(\left(n%
\left(r_{0}^{2}+r_{+}^{2}\right)^{(n-1)}-\left(n-2\right)q^{2}\,l^{2}\right)%
\left((r_{0}^{2}+r_{+}^{2})^{(n-1)}+q^{2}l^{2}\right)\Bigg.\\
&\Bigg.\left((n-2)\Xi^{2}+1\right)\Bigg)\;\Bigg\{(n-2)q^{4}\,l^{4}%
\left((3n-6)\Xi ^{2}-(n-3)\right)\Bigg. \\
&\!\!\!\Big.-2q^{2}l^{2}(r_{0}^{2}+r_{+}^{2})^{(n-1)}\left((3n-6)\Xi
^{2}-n^{2}+3\right)+n(r_{0}^{2}+r_{+}^{2})^{(2n-2)}\big((n+2)\Xi
^{2}-(n+1)\big)\Bigg\}^{-1}. \end{aligned}
\end{equation}
Figure \ref{Figure1} shows the behavior of the heat capacity as a
function of the charge parameter. It shows that $C_{\mathbf{J},Q}$
is positive in various dimensions and goes to zero as $q$
approaches its critical value (extreme black brane). Thus, the
$(n+1)$-dimensional asymptotically AdS charged rotating black
brane is locally stable in the canonical ensemble.
One may also plot $T$ versus $S$ for fixed $\mathbf{J}$ and $Q$,
and look for inflection point(s). Indeed, for systems with phase
transition the diagram has an inflection point, that demonstrates
the region of coexistence of various phases \cite{Cal}. Since
there is no an inflection point in Fig. \ref{Figure2}, the black
brane is stable.

In the grand-canonical ensemble, the stability analysis can be carried out
by calculating the determinant of Hessian matrix of the energy with respect
to $S$, $Q$ and $\mathbf{J}$. It is a matter of calculation to show that the
black brane is locally stable in the grand-canonical ensemble, since the
determinant of Hessian matrix,
\begin{equation}
H_{SQ\mathbf{J}}^{M}=\frac{\pi }{l^{2}\Xi ^{6}}\frac{\left( \frac{64}{n}%
l^{2}q^{2}+\frac{64}{n-2}(r_{0}^2+r_{+}^2)^{n-1}\right) }{((n-2)\Xi
^{2}+1)(l^{2}q^{2}+(r_{0}^2+r_{+}^2)^{n-1})(r_{0}^2+r_{+}^2)^{\frac{3}{2}n-2}%
}
\end{equation}
is positive. It is worthwhile to note that $\alpha $ has not appeared in the
thermodynamic quantities computed in this section. Thus, the charged
solutions in the Gauss-Bonnet gravity has the same thermodynamic features as
the solutions in the Einstein gravity. This fact is also true for the case
of static spherically symmetric black holes with zero curvature horizon \cite
{Cai}.

\section{Concluding Remarks}

\label{remarks}

The metric function of charged solution of Gauss-Bonnet gravity
introduced in literature is not real for the whole spacetime even
for static case \cite{Od2}. In this paper we presented the charged
rotating solutions of Gauss-Bonnet gravity which are real in the
whole spacetime and investigated the thermodynamics of them. We
found that in contrast with the case of uncharged solution, the
temperature is not equal to $f^{\prime }(r_{+})/4\pi \Xi $. The
charged solutions may be interpreted as black brane solutions with
two inner and outer event horizons for $m>m_{\mathrm{ext}}$,
extreme black holes for $m=m_{\mathrm{ext}}$ or naked singularity for $m<m_{%
\mathrm{ext}}$, where $m_{\mathrm{ext}}$ is given in Eq.
(\ref{mext}). We found that the Killing vectors are the null
generators of the event horizon, and therefore the event horizon
is a Killing horizon for the stationary solution of the
Gauss-Bonnet gravity explored in this paper. We computed the
finite action of the charged rotating solutions through the use of
counterterm method. By using the relation between the action and
free energy, we compute the conserved and thermodynamic quantities
of the solutions and showed that they satisfy the first law of
thermodynamics. Finally, we obtained a Smarr-type formula for the
mass of the black brane solution as a function of the
entropy, the charge and the angular momenta of the black brane and studied the phase behavior of the $(n+1)$%
-dimensional charged rotating black branes. We showed that there
is no Hawking-Page phase transition in spite of the charge and
angular momenta of the branes. Indeed, we calculated the heat
capacity and the determinant of the Hessian matrix of the mass
with respect to $S$, $\mathbf{J}$ and $Q$ of the black branes and
found that they are positive for all the phase space, which means
that the brane is stable for all the allowed values of the metric
parameters discussed in Sec. \ref{Therm}. It is worth to mention
that the charged rotating solutions of first order Lovelock
gravity \cite{Deh}, of second order Lovelock gravity introduced in
this paper, of third order Lovelock gravity presented in Ref.
\cite{DM} and the static solutions of continued Lovelock gravity
\cite{DBS} are stable. Thus, we conjecture that the Lovelock terms
do not have any effects on the stability of black branes with flat
horizon. This is commensurate with the fact that there is no
Hawking-Page transition for a black object whose horizon is
diffeomorphic to $\mathbb{R}^{p}$ and therefore the system is
always in the high temperature phase \cite{Wit1}. Of course this
kind of solutions in the presence of a dilaton field is not stable
even in Einstein gravity \cite{SDRP}.

Although the thermodynamics of charged rotating solutions of
Gauss-Bonnet gravity with flat horizon have been investigated in
this paper, no such a kind of solutions with spherical horizon
have been presented till now. Thus, it is worth constructing these
kinds of solutions and investigating their thermodynamics. The
charged rotating black holes with spherical horizon in Einstein
gravity are not stable in the whole phase space \cite{Cal}, and
therefore it is interesting to investigate the effects of
Gauss-Bonnet term on the thermodynamics of such kinds of
solutions.

\begin{center}
\textbf{ACKNOWLEDGMENTS}
\end{center}

This work has been supported by Research Institute for Astrophysics and
Astronomy of Maragha, Iran.


\begin{references}
\bibitem{Mal1}  J. Maldacena, Adv. Theor. Math. Phys. \textbf{2}, 231
(1998); E. Witten, \emph{ibid.} \textbf{2}, 253 (1998); O.
Aharony, S. S. Gubser, J. Maldacena, H. Ooguri, and Y. Oz, Phys.
Rep. \textbf{323}, 183 (2000).

\bibitem{Wit1}  E. Witten, Adv. Theor. Math. Phys. \textbf{2}, 505
(1998).

\bibitem{Ish} : Y. M. Cho and I. P. Neupane, Phys. Rev. D \textbf{66},
024044 (2002).

\bibitem{BDav}  N. D. Birrell and P. C. W. Davies,
\emph{Quantum Fields in Curved Space} (Cambridge University Press,
Cambridge, England, 1982).

\bibitem{Wit2}  M. B. Greens, J. H. Schwarz and E. Witten, \emph{Superstring
Theory}, (Cambridge University Press, Cambridge, England, 1987);
D. Lust and S. Theusen, \emph{Lectures on String Theory},
(Springer, Berlin, 1989); J. Polchinski, \emph{String Theory},
(Cambridge University Press, Cambridge, England, 1998).

\bibitem{Zw}  B. Zwiebach, Phys. Lett. B \textbf{156}, 315 (1985);
B. Zumino, Phys. Rep. \textbf{137}, 109 (1986).

\bibitem{Des} D. G. Boulware and S. Deser, Phys. Rev. Lett., \textbf{55}, 2656
(1985); J. T. Wheller, Nucl. Phys. B \textbf{268}, 737 (1986).

\bibitem{Cham}  A. H. Chamseddine, Phys. Lett. B \textbf{233}, 291 (1989); F.
Muller-Hoissen, Nucl. Phys. {\bf B346}, 235 (1990).

\bibitem{Lov}  D. Lovelock, J. Math. Phys. \textbf{12}, 498 (1971);
N. Deruelle and L. Farina-Busto, Phys. Rev. D \textbf{41}, 3696
(1990); G. A. MenaMarugan, \emph{ibid}. \textbf{46}, 4320 (1992);
4340 (1992).

\bibitem{BY}  J. D. Brown and J. W. York, Phys. Rev. D \textbf{47}, 1407
(1993).

\bibitem{CCM}  K. C. K. Chan, J. D. E. Creighton and R. B. Mann, Phys. Rev.
D \textbf{54}, 3892 (1996).

\bibitem{BK}  V. Balasubramanian and P. Kraus, Commun. Math. Phys. \textbf{%
208,} 413 (1999).

\bibitem{EJM}  R. Emparan, C.V. Johnson and R. C. Myers, Phys. Rev. D
\textbf{60}, 104001 (1999).

\bibitem{Deh-topologic} M. H. Dehghani, Phys. Rev. D \textbf{65}, 124002
(2002).

\bibitem{Deh} M. H. Dehghani, Phys. Rev. D \textbf{66}, 044006 (2002);
M. H. Dehghani and Khodam-Mohammadi, \emph{ibid}. \textbf{67},
084006 (2003).

\bibitem{Od1}  S. Nojiri and S. D. Odintsov, Phys. Lett. B
\textbf{521}, 87 (2001).

\bibitem{DM2}  M. H. Dehghani and R. B. Mann, Phys. Rev. D, \textbf{64},
044003 (2001); M. H. Dehghani, \emph{ibid}. \textbf{65}, 104030
(2002); M. H. Dehghani and H. KhajehAzad, Can. J. Phys.
\textbf{81}, 1363 (2003).

\bibitem{kls}  P. Kraus, F. Larsen and R. Siebelink, Nucl. Phys. \textbf{%
B563,} 259 (1999).

\bibitem{DBS}    M. H. Dehghani, N. Bostani and  A. Sheykhi, Phys. Rev.  D \textbf{73}, 104013
(2006).
\bibitem{DM}     M. H. Dehghani and  R. B. Mann,  Phys. Rev.  D \textbf{73}, 104003
(2006).

\bibitem{Myers-surf}  R. C. Myers, Phys. Rev. D \textbf{36}, 392 (1987).

\bibitem{Dav-surf}  S. C. Davis, Phys. Rev. D \textbf{67}, 024030 (2003).

\bibitem{Cal}  M. M. Caldarelli, G. Cognola, and D. Klemm, Class. Quantum
Grav. \textbf{17}, 399 (2000).

\bibitem{Haw-Ross}  S. W. Hawking and S. F. Ross, Phys. Rev. D \textbf{52}, 5865 (1995).

\bibitem{Deh4} M. H. Dehghani,  Phys. Rev. D \textbf{67},  064017
(2003).


\bibitem{Haw1+3} S. W. Hawking, Commun. Math.
Phys. \textbf{25}, 152 (1972); S. W. Hawking and G. F. R.
Ellis,\emph{The Large Scale of SpaceTime}, (Cambridge University
Press, Cambridge, England, 1973) .


\bibitem{Gub}  M. Cvetic and S. S. Gubser, J. High Energy Phys. \textbf{04},
024 (1999).

\bibitem{Beck} J. D. Beckenstein, Phys. Rev. D \textbf{7},
2333 (1973); S. W. Hawking, Nature (London) \textbf{248}, 30
(1974); G. W. Gibbons and S. W. Hawking, Phys. Rev. D \textbf{15},
2738 (1977).
\bibitem{Haw222}  C. J. Hunter, Phys. Rev. D \textbf{59}, 024009 (1998); S. W.
Hawking, C. J. Hunter and D. N. Page, \emph{ibid}. \textbf{59},
044033 (1999); R. B. Mann, \emph{ibid}. \textbf{60}, 104047
(1999); \emph{ibid}. \textbf{61}, 084013 (2000);

\bibitem{fails+1}  M. Lu and M. B. Wise, Phys. Rev. D \textbf{47},
R3095,(1993); M. Visser, \emph{ibid}. \textbf{48}, 583 (1993); T.
Jacobson and R. C. Myers, Phys. Rev. Lett. \textbf{70}, 3684
(1993); R. M. Wald, Phys. Rev. D \textbf{48}, R3427, (1993); M.
Visser, \emph{ibid}. \textbf{48}, 5697 (1993); T. Jacobson, G.
Kang and R. C. Myers, \emph{ibid}. \textbf{49}, 6587,(1994); V.
Iyer and R. M. Wald, \emph{ibid}. \textbf{50}, 846 (1994).

\bibitem{Gus} S. S. Gubser and I. Mitra, J. High Energy
Phys.,   \textbf{08}, 018 (2001).

\bibitem{Cai}  R. G. Cai and K. S. Soh, Phys. Rev. D \textbf{59}, 044013 (1999);
R. G. Cai, \emph{ibid}. \textbf{65}, 084014 (2002).

\bibitem{Od2} M. Cvetic, S. Nojiri, S. D. Odintsov, Nucl. Phys. \textbf{B628%
}, 295 (2002).

\bibitem{SDRP} A. Sheykhi, M. H. Dehghani, N. Riazi and J.
Pakravan, hep-th/0606237.

\end{references}
\end{document}